\documentstyle[12pt,epsfig]{article} 
\newlength{\dinwidth}                       
\newlength{\dinmargin}                      
\def\lsim{\mathrel{\rlap{\lower4pt\hbox{\hskip1pt$\sim$}}
    \raise1pt\hbox{$<$}}}                
\def\gsim{\mathrel{\rlap{\lower4pt\hbox{\hskip1pt$\sim$}}
    \raise1pt\hbox{$>$}}}                
\newcommand{\dk}{\Delta\kappa_{\gamma}}
\newcommand{\lam}{\lambda_{\gamma}}
\begin{document}
\vspace*{1cm}
\centerline{MPI/PhT/96--105\hfill}
\vspace*{1cm}
\begin{center}  \begin{Large} \begin{bf}
Electroweak Physics at HERA: \\
Introduction and Summary\footnote{
To appear in the {\it Proceedings of the Workshop on Future Physics at HERA}.
The complete report by the working group on electroweak physics at HERA is
available from http://www.desy.de/~heraws96/proceedings.}
\\
  \end{bf}  \end{Large}
  \vspace*{5mm}
  \begin{large}
R.\ J.\ Cashmore$^a$, E.\ Elsen$^b$, B.\ A.\ Kniehl$^{cd}$, 
H.\ Spiesberger$^c$\\ 
  \end{large}
\end{center}
$^a$ Department of Physics, University of Oxford,
     1~Keble Road, Oxford OX1~3NP, UK\\
$^b$ Deutsches Elektronen-Synchrotron DESY,
     Notkestra\ss e~85, 22603~Hamburg, Germany\\
$^c$ Max-Planck-Institut f\"ur Physik (Werner-Heisenberg-Institut),
F\"ohringer Ring~6,\\
\phantom{$^c$} 80805~Munich, Germany\\
$^d$ Institut f\"ur Theoretische Physik,
Ludwig-Maximilians-Universit\"at, Theresienstra\ss e~37,\\ 
\phantom{$^d$} 80333~Munich, Germany\\
\begin{quotation}
\noindent
{\bf Abstract:}
A high luminosity upgrade of HERA will allow the measurement of 
standard model parameters and the neutral current couplings of quarks.  
These results will have to be consistent with other precision 
measurements or indicate traces of new physics.  The analysis of $W$ 
production will complement future results of LEP~2 and the Tevatron.  
We summarize the main results and conclusions obtained by the working 
group on {\it Electroweak Physics} concerning the potential of future 
experimentation at HERA.
\end{quotation}
%
\section{Introduction}

The DESY $ep$ collider HERA is a unique place to explore the
structure of the proton, in particular at low Bjorken $x$ and large
momentum transfer $Q^2$ and at the same time to probe the theory of
electroweak interactions in the regime of large spacelike $Q^2$,
extending previous measurements at fixed target experiments by more
than two orders of magnitude. This is complementary to what can be
accessed at LEP and hadron colliders in searches for deviations
from the standard model (SM).  In the 1987 \cite{ws87} and 1991
\cite{ws91} HERA workshop proceedings and elsewhere \cite{spi},
comprehensive reviews on electroweak physics at HERA \cite{bar,buc}
and the influence of radiative corrections \cite{dba,kra} have been
published. In the meantime HERA has started and proved to work
reliably. The detectors have shown to operate successfully and to
be able to stand the special environment of an $ep$ machine. 

The experience collected in the first years of experimentation at HERA
allowed us to consider in the present workshop in some detail
experimental problems, like systematic uncertainties due to the energy
scale, the luminosity measurement or the measurement of the
polarization, as well as limited acceptance and efficiencies. In this
respect, the contributions to this workshop go beyond earlier studies.

The principal goal of the present workshop, on {\it Future Physics at
  HERA}, is to explore the physics potential attainable by possible
machine and detector upgrades with respect to the various options
under discussion:
(i) high (1~TeV) versus design (820~GeV) proton energy;
(ii) fixed-target versus collider mode;
(iii) light or heavy nuclei versus protons;
(iv) polarized versus unpolarized protons;
(v) polarized versus unpolarized electrons and positrons; and
(vi) high (1~fb$^{-1}$) versus design (250~pb$^{-1}$) luminosity.
Working group 2 has analyzed these options and concentrated on the 
interesting cases for {\it Electroweak Physics}.  In this introductory 
report, we shall summarize the main results obtained by the various 
subgroups and draw conclusions.

Prior to presenting an overview of the various subgroup activities and
reporting the key results, we preselect from the upgrade options
enumerated above those which will prove most useful for the study of
electroweak physics at HERA, and argue why the residual options will
have marginal advantage or even disadvantage.  In order to suppress
the impact of the (well-tested) electromagnetic interaction in
neutral-current (NC) deep inelastic scattering and to gain sensitivity
to the $W$-boson mass in charged-current (CC) deep inelastic
scattering, large values of $Q^2$ and thus centre-of-mass energy
$\sqrt s$ are required.  Increasing the proton energy $E_p$ by 22\%
while keeping the lepton energy $E_e$ fixed, as in option~(i), will
only increase $\sqrt s$ by 10\%, and will insignificantly improve the
electroweak-physics potential.  By the same token, the fixed-target
mode of option~(ii) will reduce the centre-of-mass energy to $\sqrt
s=7.6$~GeV, assuming the design lepton energy of $E_e=30$~GeV, and
will so render the study of electroweak physics much more difficult.
Clearly, in order to perform precision tests of the electroweak
theory, we will need as much luminosity per experimental setup as
possible, so that option~(vi) must receive high priority.  Also, to
disentangle the helicity structure of the weak NC and, in particular,
to measure the vector and axial-vector couplings of quarks to the
$Z$-boson, beams of longitudinally polarized electrons and positrons
must be available with appropriate luminosities \cite{zet}, as in
option~(v).  On the other hand, options~(iii) and (iv) are not useful
for our purposes, since the structure functions of nuclei and
polarized protons are at present poorly known, which would jeopardize
electroweak precision tests.  In addition, the total available
luminosity would be distributed among too many different experimental
setups and probably decreased due to the additional construction
periods. In conclusion, options~(v) and (vi) will be crucial for 
electroweak studies.

At HERA, investigations of electroweak physics may be classified
according to two categories of processes: First, there is the more
conventional measurement of inclusive deep inelastic scattering. Due
to reasonably high cross sections, not only the measurement of total
cross sections and their ratios, but also of differential cross
sections, will allow us to envisage precision tests of the electroweak
standard model. Experiments may be interpreted in terms of a
measurement of the basic standard model parameters, e.g.\ the mass of
the $W$ boson, $m_W$, or the top-quark mass, $m_t$. The potential of
HERA for this kind of analysis was investigated in Ref.\ \cite{prec96}
and is summarized in section 2.1.  Confronting within the standard
model measurements obtained at HERA with results from other
experiments will constitute one important test of our present
understanding of the electroweak interactions.  Another type of test
of the standard model is possible by measuring quantities which are
not free parameters in the standard model Lagrangian and comparing the
experimental results with corresponding theoretical predictions.  In
particular, the measurement of NC couplings of quarks is a test of
this kind and was studied in Ref.\ \cite{nccoupl} (section 2.2).  More
general quantities generalizing the standard model Lagrangian, like
the $\rho$-parameter or the $S$, $T$, $U$ parameters have been
considered earlier in Ref.\ \cite{spi}.  Finally, an analysis aiming
to assess the sensitivity for additional heavy charged gauge bosons
$W'$ (section 2.4) starts to overlap with the activities of the
working group {\it Beyond the Standard Model}.

The second class of processes with a potential to study electroweak
physics comprises scattering processes into exclusive final states.
The aim to measure processes like Higgs-boson production (see section
2.7), the production of $b$-quarks or, most interestingly, of $W$ and
$Z$ bosons (section 2.5), and to compare corresponding experimental
results with the predictions of the standard model is an experimental
challenge by itself. In the latter case, in order to quantify the
results of such measurements testing the validity of the standard
model in the gauge sector, it has become customary to generalize the
standard model Lagrangian by introducing non-standard, so-called
anomalous, couplings.  Eventually, the information obtained by
studying these processes will be concentrated in statements about
these anomalous 3-boson couplings, $\Delta\kappa_V$ and $\lambda_V$
for $WWV$ ($V=\gamma$, $Z$) and $h_i^V$ for $Z\gamma V$.

Theoretical uncertainties due to an incomplete knowledge of the
structure function input, as well as unknown higher-order QCD
corrections and uncertainties in the scale of $\alpha_S$, deserve
particular attention since they could severely limit the usefulness of
electroweak physics analyses of deep inelastic scattering 
\cite{prec96,bakqcd}. Each of the
subgroups has therefore undertaken particular efforts to demonstrate
that already with the present knowledge about structure functions
sensible measurements can indeed be performed. With high luminosity
available, future measurements at HERA are bound to improve the
situation.

\section{Summaries of the individual contributions}

\subsection{Electroweak precision tests}

The most obvious question about the possible contribution of HERA for
electroweak physics tests is to which extent the basic standard model
parameters $m_W$, $m_t$ and $m_H$ can be constrained by precision
measurements of deep inelastic scattering cross sections. From earlier
work it is known that measurements at HERA without any additional
input from other experiments are very similar to a measurement of
$G_{\mu}$, the $\mu$ decay constant, but at $\langle Q^2 \rangle
=O(3000\,$GeV$^2$).

\begin{figure}[tbhp]\centering
\mbox{\epsfig{figure=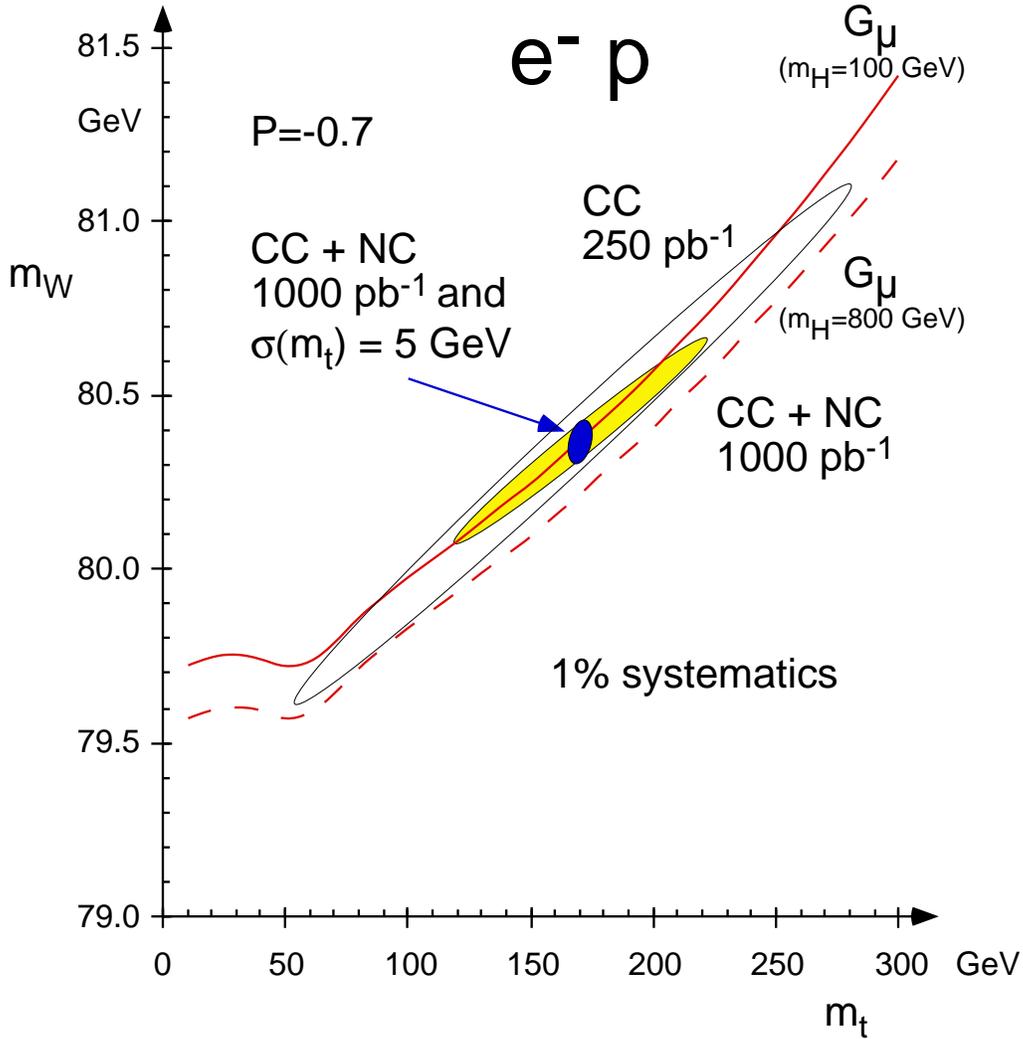,width=0.80\textwidth}}
\caption{\it 1$\sigma$-confidence contours in the ($m_W,m_t$) plane
  from polarized electron scattering ($P=-0.7$), utilizing charged
  current scattering at HERA alone with an integrated luminosity of
  250 pb$^{-1}$ (large ellipse), neutral and charged current scattering
  at HERA with 1000 pb$^{-1}$ (shaded ellipse), and the combination of
  the latter HERA measurements with a direct top mass measurement with
  precision $\sigma(m_t) = 5\,$GeV (full ellipse). The $m_W$-$m_t$
  relation following from the $G_{\mu}$-constraint is also shown for
  two values of $m_H$.}
\label{fig_nccc_sens}
\end{figure}

In the report of the subgroup on {\it Electroweak Precision Tests}, it
is pointed out that, although an interpretation of deep inelastic
scattering measurements in terms of either $m_W$ or $m_t$ results in
rather large errors, HERA data will put a rather stringent constraint
on the interrelation of these two parameters.  Fig.\ 
\ref{fig_nccc_sens} presents results for a measurement with 1000
pb$^{-1}$ of data from polarized neutral and charged current electron
proton scattering.  The corresponding 2$\sigma$-contour is represented
by the shaded ellipse. Projecting it onto the axes results in
precisions of $\delta m_t = \pm 50\,$GeV and $\delta m_W = \pm
290\,$MeV. These values are more than a factor of 2 better than what
can be obtained from charged current scattering alone with the smaller
luminosity of 250 pb$^{-1}$ (see the large ellipse in the figure).
Anticipating future direct high precision measurements of $m_t$ and
$m_W$, a comparison with these data will provide a stringent test of
the standard model. One way to quantify the measurement is to combine
HERA data on neutral and charged current data with a direct top mass
measurement of $\pm 5\,$GeV.  Such a test yields $\delta m_W = \pm
60\,$MeV. This scenario assumes a value of $1\,\%$ relative systematic
uncertainty, which represents a serious experimental challenge. The
figure also shows the relation between $m_W$ and $m_t$ following from
the $G_{\mu}$ constraint. The upper full line is for a Higgs mass of
$m_H = 100\,$GeV, the lower dashed one for $m_H = 800\,$GeV. Note that
the confidence ellipses derived from HERA measurements are also
obtained for a fixed Higgs mass value which was chosen to be
$100\,$GeV. They would be shifted downwards much the same as the lines
describing the $G_{\mu}$ constraint for a Higgs mass of $m_H =
800\,$GeV. It is obvious from this figure that with such a high
precision one would be able to constrain the allowed range of Higgs
masses provided one has available a second precise measurement of the
$W$ boson mass.

Rather than taking into account a more or less well-motivated fixed 
size of experimental systematic errors, the {\it Electroweak Precision 
Tests} subgroup decided to investigate the dependence of the 
measurement error on $m_W$ on the systematic uncertainty in a 
range up to conservative values of $5\,\%$.  Future experiments are 
expected to reduce this value of systematic uncertainty, as well as 
uncertainties from parton distribution functions, considerably.

Comparing different scenarios of beam setups (electrons versus 
positrons and degree of longitudinal polarization), it turned out that 
experiments with left-handed electrons alone would give the highest 
precision from both NC and CC scattering.  This is essentially a 
consequence of the need to have as much data as possible and 
thus the process with the highest cross section is preferred.

\subsection{NC couplings of quarks}

\begin{figure}[tbhp] \begin{center}
\mbox{\epsfig{figure=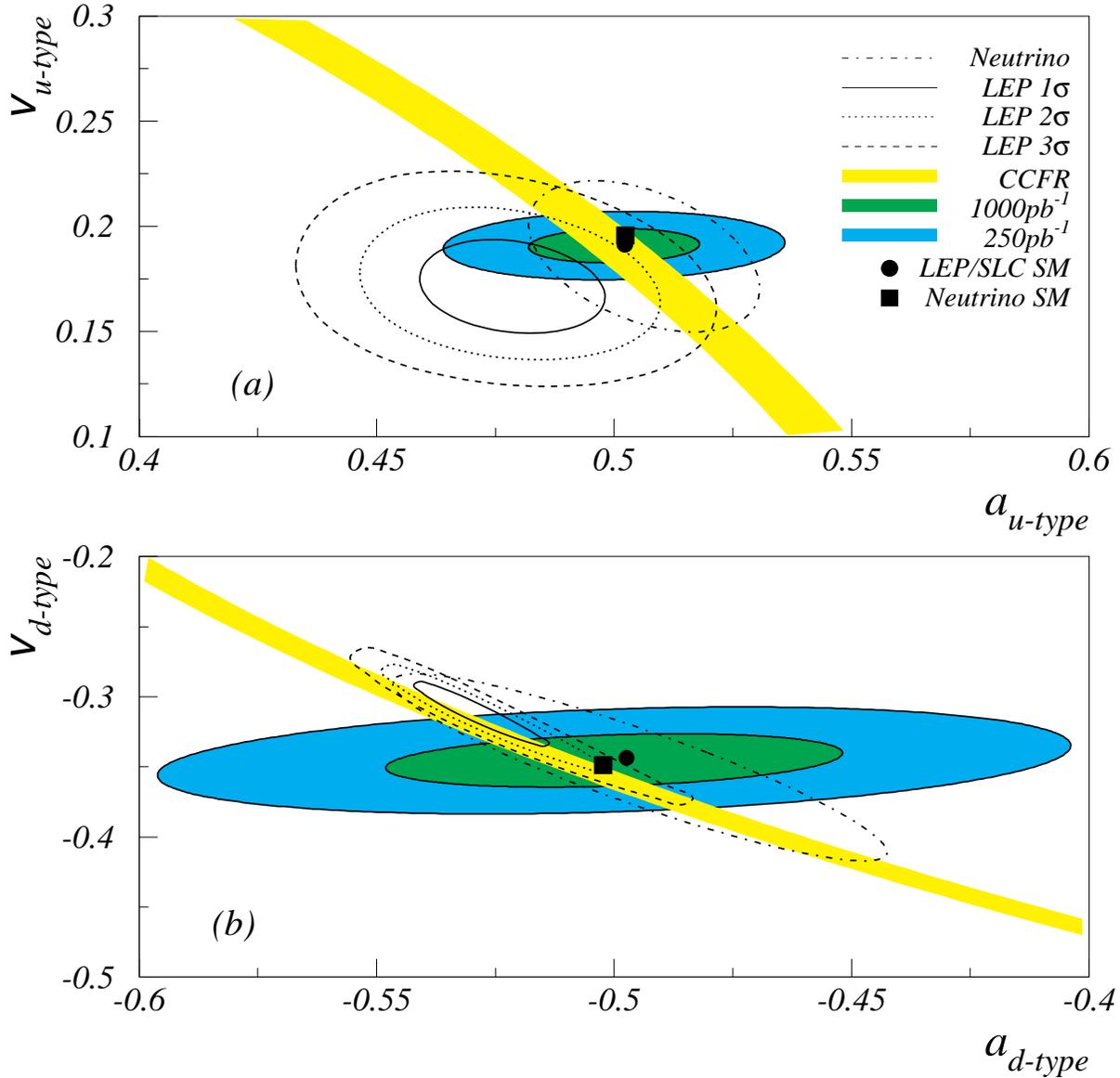,width=0.95\textwidth}}
\caption{\it 
  Summary of measurements of $u$-type (a) and $d$-type (b) quark
  couplings to the $Z^0$.  The results of a measurement at HERA are
  shown as the shaded ellipses.  The outer ellipse shows the result
  which would be obtained with 250~pb$^{-1}$ divided equally between
  the four lepton beam charge/polarization combinations, the inner
  ellipse shows the result which would be obtained with 1000~pb$^{-1}$
  equally divided.  Fits for the couplings of the $u$ ($d$) quarks
  were performed with the $d$ ($u$) quark couplings fixed at their SM
  values.  The open ellipse drawn with a dash-dotted line shows the
  one standard deviation ($1\sigma$) contour obtained in a fit of the
  four chiral couplings of the $u$ and $d$ quarks to a compilation of
  neutrino DIS data. The solid, dotted and dashed ellipses show the 1,
  2 and 3 $\sigma$ limits of the combined LEP/SLD results for the $c$
  and $b$ quark couplings. The shaded band shows the result obtained
  by the CCFR collaboration from the ratio of NC and CC cross sections
  projected onto the quark coupling plane.  In the case of the CCFR
  the couplings of the $u$ ($d$) quarks are obtained with the $d$
  ($u$) quark couplings fixed at their SM values. SM coupling values
  including radiative corrections appropriate for comparison with
  $e^+e^-$ (circle) and neutrino (square) measurements are also
  shown.}
\label{fignccoupl} \end{center} 
\end{figure}

Measurements of NC and CC cross sections with longitudinally polarized 
electrons and positrons would provide enough information to 
disentangle the neutral current couplings of the light $u$ and $d$ 
quarks.  This is demonstrated in the contribution of the subgroup on 
{\it Measurement of NC Couplings}.  The analyses are based on the NC 
and CC cross sections.  In a scenario with $1000\,{\rm pb}^{-1}$ 
divided equally between the four charge/polarization combinations, all 
four $u$ and $d$-type vector and axial-vector couplings can be 
measured with resulting fractional errors on $a_u$, $v_u$, $a_d$ and 
$v_d$ of $6\,\%$, $13\,\%$, $17\,\%$ and $17\,\%$, respectively (see 
Fig.\ \ref{fignccoupl}).  Even higher precisions could be achieved by 
constraining two of the couplings to their standard model values.  
These results are comparable with the heavy quark couplings determined 
at LEP~1.

These studies are based on full Monte Carlo simulations taking into
account the present knowledge of the ZEUS detector performance and the
current analysis methods.  Future upgrades to the detector and
improvements in the understanding of calibration of existing detector
components will serve to improve the precision of the measurements. In
addition, a detailed investigation of uncertainties from parton
distribution functions, entering into the analyses basically via
ratios of $u$ and $d$-type quark densities, and a comparison with the
experimental errors have shown that this latter source of
uncertainty will not be the limiting factor in determining NC
couplings of quarks at HERA.

In the light of the intriguing $R_b$ anomaly which has been presented 
by the four LEP experiments \cite{lepewwg}, the question was raised 
whether HERA would be able to provide complementary information.  From 
earlier workshops it is known that the total rate of $b$-quark 
production is not at all small.  However, the contribution due to 
photon-$Z$ interference and pure $Z$-exchange is tiny (order 50 events 
for $100$\,pb${}^{-1}$) \cite{schuler} and much too small to be 
helpful for electroweak physics.

\subsection{$W^\prime$}

The potential of HERA for the discovery of additional heavy neutral or 
charged gauge bosons had been studied carefully in earlier workshops.  
In Ref.\ \cite{cornet} it has been pointed out that charge and 
polarization asymmetries of NC cross sections are particularly suited 
for the search and the identification of a heavy $Z'$ boson and 
exclusion limits for its mass have been given there.  A model 
independent analysis, considering the 6-dimensional space of 
$Z'f\bar{f}$ couplings was performed in Ref.\ \cite{martyn}.

\begin{figure}[tbhp]\centering
\mbox{\epsfig{figure=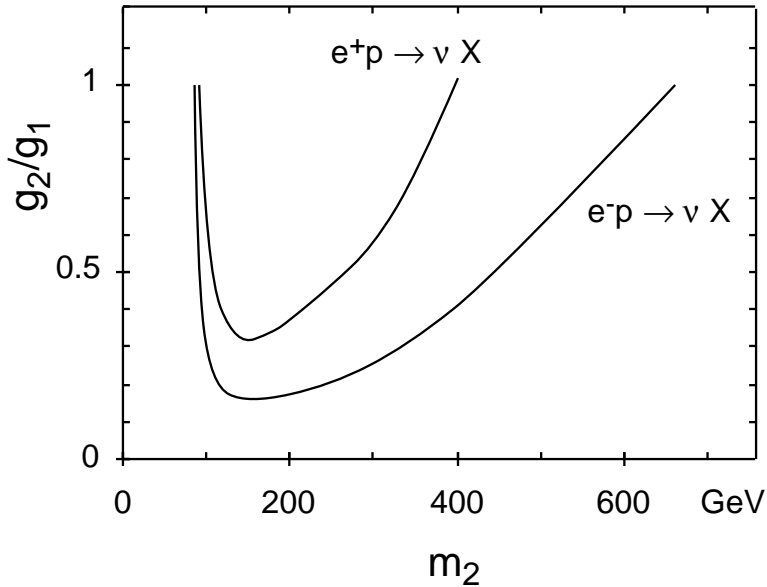,width=0.60\textwidth}}
\caption{\it 95 \% exclusion limits for a heavy $W'$ in the ($g_2/g_1$,
  $m_2$) plane.}
\label{figwprime}
\end{figure}

Similar discovery limits for a heavy charged boson $W'$ can be derived
by considering a possible deviation of the measured CC cross section
from its standard model prediction assuming the existence of a heavy
$W'$.  A possible signal could show up in a comparison of the CC cross
section at HERA with data at zero-momentum transfer, e.g.\ the
$\mu$-decay constant.  The enhancement of the CC cross section with
respect to the SM prediction in the presence of two charged $W$ bosons
is approximately given by
\begin{equation}
\frac{d\sigma(W+W')}{d\sigma_{SM}} = 1 +
\frac{2x^2}{1 + x^2 \frac{m_W^2}{m_2^2}}
\left(1-\frac{m_W^2}{m_2^2}\right) \frac{Q^2}{Q^2+m_2^2}, 
\end{equation}
where $x=g_2/g_1$ is the ratio of the coupling constants of the two 
charged bosons.  The mass of the lighter one has been identified with 
that of the standard model $W$, while the heavier one is denoted by 
$m_2$.  The exclusion limits in the ($x$, $m_2$) plane derived from 
the condition that the enhancement be larger than the statistical 
precision of the measurement of the CC cross section are shown in 
Fig.\ \ref{figwprime}.  Assuming equal coupling strengths, $g_2/g_1 = 
1$, the resulting limits are $m_2 \lsim 400$ GeV for positron 
scattering and $m_2 \lsim 630$ GeV \cite{haidt}.  These mass limits do not 
supersede corresponding limits from the Tevatron and we have not 
considered them in further detail.

\subsection{$W$ production}

The study of single $W$ production offers the challenging opportunity
to test the nonabelian structure of the standard model at HERA, in
particular to search for deviations of the $WW\gamma$ couplings from
its standard model values \cite{gaemers,baur}. In a new study of this
process during the present workshop \cite{val}, a discussion of the
event topology and kinematical cuts to optimize the event selection is
presented. It is shown that HERA offers greater sensitivity to
anomalous values of $\Delta \kappa_{\gamma}$ than $\lambda_{\gamma}$
and therefore complements measurements made at the Tevatron and LEP~2
where the sensitivity to $\lambda_V$ is greater. Since single $W$
production at HERA is quite insensitive to anomalous $WWZ$ couplings,
unlike $WW$ production at the Tevatron or LEP~2, measurements at HERA
will be important to identify the nature of possible deviations, if
they would be observed.

\begin{figure}[tbhp] 
\centerline{\psfig{file=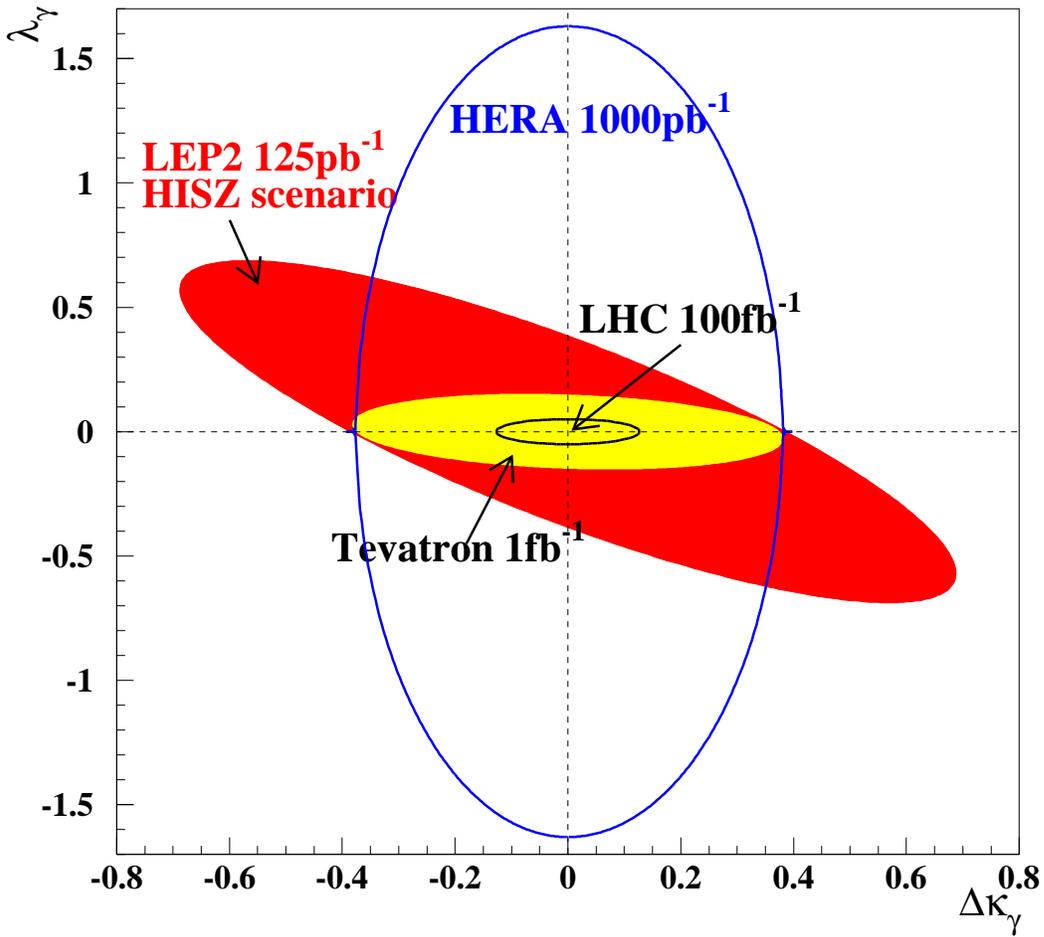,height=15cm}}
\caption{\it 
  Projected 95\% confidence level sensitivity limits for $WW\gamma$
  couplings determined from the single $W$ production cross section at
  HERA and from $W\gamma$ production at the Tevatron and the LHC.  The
  solid shading indicates the limits for the $WWV,V=\gamma,Z$
  couplings from $WW$ production at LEP~2 assuming the HISZ scenario.}
\label{figure:expts}
\end{figure}

The sensitivity to anomalous $WW\gamma$ couplings has been studied for
various integrated luminosities and at two center-of-mass energies.
The resulting $95\,\%$ confidence level limits at $\sqrt{s} = 300$ GeV
are given in Table \ref{tlimits}. At $\sqrt{s}=346$ GeV, the limits
for $\int{\cal L}dt = 1000\,{\rm pb}^{-1}$ are $-0.27 <
\Delta\kappa_{\gamma} < 0.26$ and $-1.26 < \lambda_{\gamma} < 1.28$.
They are limited by statistical rather than systematic errors.  For 
$\int{\cal L}dt=1000$ pb${}^{-1}$ the future sensitivity on anomalous 
values of $\Delta\kappa_{\gamma}$ which can be obtained at HERA is 
competitive with projected limits from $W\gamma$ production at the 
Tevatron (see Fig.\ \ref{figure:expts}).  The bounds from LEP~2 shown 
in this figure are based on the auxiliary assumption that the $WWZ$ 
and $WW\gamma$ couplings are not independent from each other (HISZ 
scenario) and can thus not be compared with the HERA results on equal 
footing.

\begin{table}[tbh]
\begin{center}
\begin{tabular}{|r |c |c|} \hline
$\int{\cal L}dt$     &      $\dk$       &     $\lam$  \\ \hline\hline
$100~{\rm pb}^{-1}$  & $-1.43<\dk<0.95$ & $-2.93<\lam<2.94$ \\ \hline
$200~{\rm pb}^{-1}$  & $-0.87<\dk<0.72$ & $-2.46<\lam<2.47$ \\ \hline
$1000~{\rm pb}^{-1}$ & $-0.38<\dk<0.38$ & $-1.65<\lam<1.66$ \\ \hline
\end{tabular}
\end{center}
\caption{\it 
  95\% CL limits derived for $WW\gamma$ couplings from the measurement
  of $\sigma(ep \rightarrow eWX)$ at HERA for the nominal
  center-of-mass energy of 300 GeV\label{tlimits}.}
\end{table}

\subsection{Radiative NC scattering}

Radiative deep inelastic scattering offers another
possibility to study trilinear gauge boson couplings. Radiative CC
scattering and its potential to probe the $WW\gamma$ couplings has
been studied in Ref.\ \cite{helbig}. As a new contribution to this
workshop, Ref.\ \cite{illana} investigated whether $Z\gamma\gamma$
couplings can be tested in radiative NC scattering. Contributions due
to $ZZ\gamma$ couplings are suppressed by two $Z$ propagators. Since
the rates are too small to exploit differential cross sections,
estimates for bounds are obtained from total cross sections taking
into account realistic cuts to improve the sensitivity to this source
of new physics.  HERA will explore these couplings in a different
kinematic regime than at LEP~2, NLC or hadron colliders. However, it 
turns out that
competitive bounds on these anomalous couplings cannot be achieved,
even with the future luminosity upgrades.

\subsection{SM Higgs-boson production}

The prospects of producing light SM Higgs bosons at HERA under nominal
conditions were discussed in the 1987 proceedings \cite{gae}.  In the
meantime, LEP has ruled out the mass range $M_H<65.2$~GeV at the 95\%
confidence level \cite{gri}.  An update of the 1987 analysis assessing the
benefits from luminosity and proton energy upgrades may be found in
Ref.\ \cite{bak}.  $W^+W^-$ and $ZZ$ fusion are by far the most
copious sources of SM Higgs bosons at HERA.  In the mass range
65~GeV${}<M_H<{}$100~GeV, the Higgs boson decays with about 90\%
branching fraction to $b\bar b$ pairs, so that its visibility at HERA
will suffer from severe intrinsic backgrounds due to continuum $b\bar
b$ production.  If such a Higgs boson exists, its production cross
section at HERA will be below 6 (9)~fb for $E_p=820$~GeV (1~TeV).  It
is therefore unlikely that a signal can be established in the $b\bar
b$ channel, even if the luminosity and/or proton energy are upgraded
leaving this terrain to LEP~2.

\section{Conclusions}

The observation of the propagator effect due to $W$ exchange was one 
of the first results from HERA at high $Q^{2}$ after decades of 
searching for deviations of the linearly rising cross section of the 
CC process.  In the meantime, the finite mass of the $W$ boson 
responsible for this effect, the propagator mass, has been measured 
with an accuracy of a few GeV at HERA.  First candidate events for the 
direct production of the $W$ boson have also been observed at HERA.  
In order to turn these observations into measurements of parameters 
and conclusive tests of the standard model electroweak sector the 
luminosity has to be increased tremendously.  Luminosities of $1\,{\rm 
fb}^{-1}$ and polarized beams will make these measurements possible.

The cross section for the production of $W$ bosons is the order of 
$1\,$pb so that it is an experimental challenge to establish the 
experimental signal and it is essential to consider all available 
decay channels.  The study performed in this report shows that already 
with an integrated luminosity of $250\,{\rm pb}^{-1}$ deviations from 
the standard model couplings parametrized in terms of 
$\Delta\kappa_{\gamma}$ and $\lambda_{\gamma}$ can be tested at a 
level which is comparable to present collider results.

While for $W$ production neither lepton charge nor polarization is a 
prerequisite, the precision measurement with the charged current 
process profits especially from $e^-p$ scattering due to the roughly 
threefold larger cross section as compared to $e^+p$ scattering.  For 
an analysis in terms of NC quark couplings it is indispensable to have 
available both charge states, while not necessarily with equal 
luminosity.  To disentangle up- and down-type quark vector and 
axial-vector couplings beams have to be polarized.  A proper choice of 
polarization would also render the NC data useful, and evidently 
enhance the significance of the CC data, for the precise 
determination of e.g. $m_{W}$ or $m_{t}$.

The result of these measurements constitutes an important test of the 
standard model by comparing precision measurements of SM parameters 
obtained at HERA with those at other experiments.  Differences 
appearing in such tests have always stimulated extensive research.  
The separation of the light up- and down-type quark couplings at HERA 
would complement the achievements of LEP in the heavy quark sector.


\end{document}